\documentclass[preprint,aps,prb,showpacs,a4paper,superscriptaddress]{revtex4}

\usepackage{graphicx}
\usepackage{amsmath}
\begin{document}
\title{Role of pseudospin in quasiparticle interferences in epitaxial graphene, \\ 
probed by high resolution scanning tunneling microscopy}

\author{P.~Mallet}
\email[electronic address: ]{pierre.mallet@grenoble.cnrs.fr}
\altaffiliation{The two first authors contributed equally to this work.}
\affiliation{Institut N\'eel, CNRS-UJF, BP166, 38042 Grenoble, France\\}

\author{I.~Brihuega}
\affiliation{Max-Planck-Institut f\"{u}r Festk\"{o}rperforschung, Heisenbergstrasse 1, D-70569 Stuttgart, Germany\\}
\affiliation{Departamentao de F\'{i}sica de la Materia Condensada, Universidad Autonoma de Madrid, E-28049 Madrid, Spain\\}

\author{S.~Bose}
\affiliation{Max-Planck-Institut f\"{u}r Festk\"{o}rperforschung, Heisenbergstrasse 1, D-70569 Stuttgart, Germany\\}
\affiliation{Center for Excellence in Basic Sciences, University of Mumbai, Vidhyanagari Campus, Mumbai-400098, India\\}

\author{M.M.~Ugeda}
\affiliation{Departamentao de F\'{i}sica de la Materia Condensada, Universidad Autonoma de Madrid, E-28049 Madrid, Spain\\}

\author{J.M. G\'{o}mez-Rodr\'{i}guez}
\affiliation{Departamentao de F\'{i}sica de la Materia Condensada, Universidad Autonoma de Madrid, E-28049 Madrid, Spain\\}

\author{K.~Kern}
\affiliation{Max-Planck-Institut f\"{u}r Festk\"{o}rperforschung, Heisenbergstrasse 1, D-70569 Stuttgart, Germany\\}
\affiliation{Institut de Physique des Nanostructures, Ecole Polytechnique F\'ed\'erale de Lausanne, CH-1015 Lausanne, Switzerland\\}

\author{J.~Y.~Veuillen}
\affiliation{Institut N\'eel, CNRS-UJF, BP166, 38042 Grenoble, France\\}

\date{\today}

\begin{abstract}

Pseudospin, an additional degree of freedom emerging in graphene as a direct consequence of its honeycomb atomic structure, is responsible of many of the exceptional electronic properties found in this material. This article is devoted to provide a clear understanding of how such graphene's pseudospin impacts the quasiparticle interferences of monolayer (ML) and bilayer (BL) graphene measured by low temperature scanning tunneling microscopy and spectroscopy. We have used this technique to map, with very high energy and space resolution, the spatial modulations of the local density of states of ML and BL graphene epitaxialy grown on SiC(0001), in presence of native disorder. For the first time, we perform a Fourier transform analysis of such modulations including wavevectors up to unit-vectors of the reciprocal lattice. Our data demonstrate that the quasiparticle interferences associated to some particular scattering processes are suppressed in ML graphene, but not in BL graphene. Most importantly, interferences with $2q_F$ wavevector associated to intravalley backscattering are not measured in ML graphene, even on the images with highest resolution where the graphene honeycomb pattern is clearly resolved. In order to clarify the role of the pseudospin on the quasiparticle interferences, we use a simple model which nicely captures the main features observed on our data. The model unambiguously shows that graphene's pseudospin is responsible for such suppression of quasiparticle interferences features in ML graphene, in particular for those with $2q_F$ wavevector. It also confirms scanning tunneling microscopy as a unique technique to probe the pseudospin in graphene samples in real space with nanometer precision. Finally, we show that such observations are robust with energy and obtain with great accuracy the dispersion of the $ \pi$ bands for both ML and BL graphene in the vicinity of the Fermi level, extracting their main tight binding parameters.

\end{abstract}





\maketitle

\section{Introduction}

Graphene is a very unique two-dimensional system, hosting quasiparticles which behave as massless Dirac fermions   \cite{Wallace 1947,Castro 2009}. Indeed, at low energy, they show a linear and isotropic dispersion relation, at the two opposite points (valleys) $K$ and $K'$ of the first Brillouin zone \cite{Wallace 1947,Castro 2009}. This behavior is a consequence of the honeycomb structure of the graphene lattice: The quasiparticle wavefunctions are built on two unequivalent $A$ and $B$ triangular sublattices of carbon atoms, which introduces a new degree of freedom, the pseudospin. The pseudospin is defined by the phase relation existing between the two sublattice components of the wavefunctions. Such phase relation is intimately tied to the direction of the quasiparticle momentum:  In monolayer (ML) graphene, the pseudospin is either parallel or antiparallel to the momentum, which leads to chiral Dirac fermions  \cite{Castro 2009, Katsnelson  2006}.

The pseudospin and the related electronic chirality have a key impact on the low energy band structure, and eventually on the electronic transport properties in graphene. The most striking one is the chiral half-integer quantum Hall effect measured at high magnetic field reported in 2005  \cite{Novoselov 2005, Zhang 2005}. At zero or low magnetic field, the pseudospin also impacts the electronic transport properties. Indeed, as predicted in the pioneering theoretical work of Ando et al., the pseudospin prevents backscattering processes in ML graphene in presence of long range disorder   \cite{Ando 1998,Ando 2005}. This has measurable consequences such as weak antilocalisation phenomena  \cite{McCann 2006, Falko 2007, Wu 2007, Tikhonenko 2009} and Klein tunneling   \cite{Katsnelson 2006, Young&Kim 2009, Stander 2009}. 
However, such zero or low field transport properties do not show up readily in graphene samples which contain a certain amount of atomic size impurities (substitional defects, vacancies), which generate additional intervalley scattering processes ) \cite{Ando 2005, McCann 2006, Falko 2007}. Such localized defects dramatically affect the electronic mobility  \cite{Chen 2009}  and prevent the observation of weak antilocalisation  \cite{Tikhonenko 2008}.

In close relation with transport measurements, the impact of point defect scatterers upon the local density of states (LDOS) of graphene is a central issue. From a theoretical point of view, both the intravalley and intervalley scattering processes are likely to reflect in LDOS modulations associated to Friedel charge density oscillations generated by the defects  \cite{Friedel 1958}. Recently, theoretical works were reported by several groups. They use the Green's function formalism, in order to compute the LDOS modulations due to a single atomically-sharp impurity  \cite{Mariani 2007, Bena 2008, PeregBarnea 2008, Bena 2009, Peres 2009, Bacsi 2010}. Such publications are triggered by the possible direct comparison of the theoretical results with experimental data obtained by scanning tunneling microscopy and spectroscopy, a technique well suited for probing the surface LDOS modulations at the atomic scale  \cite{Crommie 1993, Hasegawa 1993}.  

A critical issue pointed out in the above theoretical papers  is the possible existence of long-range LDOS modulations of wavevector $2q_{F}$  associated to intravalley backscattering off atomically-sharp defects (which break the $AB$ sublattice symmetry). A consensus seems to be achieved: Such LDOS modulations are present on each sublattice, but with opposite phase between them. Thus the two contributions cancel each other when averaged on the lattice unit cell. As a result, the amplitude of the $2q_{F}$ coarse-grained LDOS modulations around point defects in graphene is strongly reduced, with a $1/r^2$ decay instead of the standard $1/r$ decay found in conventional two-dimensional (2D) systems  \cite{CheianovFalko 2006, Mariani 2007, Bena 2008, PeregBarnea 2008, Bena 2009, Peres 2009, Bacsi 2010}. Recently, it was theoretically shown that such LDOS modulations can be enhanced using confined geometries, such as elliptic quantum corrals, which favor multiple scattering processes and restores backscattering  \cite{Vaishnav 2011}.

To our knowledge, only a few experimental STM studies devoted to quasiparticle interferences in graphene has been reported so far. Most of the work is related to epitaxial graphene on SiC, and focuses on the  ($\sqrt{3}\times\sqrt{3})R30^{\circ}$  superstructure surrounding the surface impurities or close to armchair edges, which is associated to intervalley scattering processes  \cite{Mallet 2007, Rutter 2007, Brihuega 2008, Simon 2009, Dujardin 2010, Ye 2010, Mahmood 2012}. This  ($\sqrt{3}\times\sqrt{3})R30^{\circ}$   pattern is also routinely measured in highly oriented pyrolitic graphite (HOPG)  \cite{MizesFoster 1989, Ruffieux 2000, Ruffieux 2005, Fukuyama 2002, Kobayashi 2005, Sakai 2010, Ugeda 2010} and has been recently observed in weakly coupled graphene on metals  \cite{Ugeda 2011, Martinez 2011}. It has been used to derive the dispersion relation on bilayer (BL)  graphene  \cite{Rutter 2007, Simon 2009}, and more temptatively on ML graphene on SiC(0001)   \cite{Rutter 2007}. In Ref. [  \cite{Brihuega 2008}], some of us have shown that the  ($\sqrt{3}\times\sqrt{3})R30^{\circ}$  pattern measured on ML presents dissimilarities with respect to BL, which are ascribed to graphene's pseudospin. 

More difficult to capture are the long range LDOS modulations associated to intravalley scattering processes in graphene. In 2007, J. Stroscio's group reported such long range modulations with wavevector $2q_{F}$ in BL graphene on SiC(0001)  \cite{Rutter 2007}. In 2008, we have shown that such $2q_{F}$ modulations are indeed present in BL graphene, but that they are lacking in ML graphene  \cite{Brihuega 2008}, in agreement with the above theoretical predictions.  To our surprise, very well pronounced long range LDOS modulations were reported in exfoliated ML graphene on SiO$_{2}$  by M. Crommie's group  \cite{Zhang 2009}. This is a puzzling result, which hardly matches with the strongly attenuated $2q_{F}$  LDOS modulations predicted by theory, as stated by the authors themselves. Very recently, interesting results have been found on ML graphene deposited on boron nitride: Long range LDOS modulations have been detected close to a smooth graphene step edge, with a strong decay with respect to the one expected for systems without pseudospin   \cite{Xue 2012}. Quantitative analysis is made possible thanks to a lateral averaging of the LDOS modulation in the direction parallel to the step. The results have been found consistent with the electronic chirality of graphene.

 To shed light on such possible observation in the case of atomic sharp impurities, we present in this paper new STM data obtained on terraces of ML and BL graphene on SiC(0001), with terrace dimensions larger than 100 nm. The spectroscopic LDOS maps of these systems are measured by STM at 5K, and are analysed by 2D Fourier transform (FT). Our main motivation is to perform a complete analysis of the features present in the FT-LDOS maps of ML and BL graphene, and to highlight how the pseudospin impacts the quasiparticle interferences in ML graphene. Importantly, this study confirms, with a higher accuracy than in Ref. [  \cite{Brihuega 2008}], that the  $2q_{F}$ modulations are not detected in ML graphene on SiC(0001), despite the presence of atomically-sharp impurities, and offers a very intuitive explanation based on the role of the pseudospin. In addition,  we performed the analysis of FT-LDOS maps as a function of energy in the vicinity of the Fermi level, which allows us to extract the detailed low-energy dispersion relation both for ML and BL graphene terraces grown on the same SiC(0001) substrate.

The structure of the paper is the following: we give in section II the experimental methods, and in particular we explain the caution needed to get highly-resolved STM data in $k$ space. In section III, we discuss the general features that should show up in the FT-LDOS maps of graphene in the absence of pseudospin, based on standard Fermi surface and joint density of states (JDOS) considerations.  Section IV is devoted to the STM measurements achieved on ML graphene. We demonstrate that the FT-LDOS maps lack the central ring of radius  $2q_{F}$ associated to intravalley backscattering, and show that replica of this ring are found around the first-order graphene lattice spots. We also show the split-ring features related to intervalley scattering, with unprecedented $k$ resolution. We extract from these features the quasiparticle dispersion relation of ML graphene on SiC(0001).
In section V, we introduce a simple model to understand why the FT-LDOS map measured on ML differs from the one expected in section III. We use single particle scattering considerations within the tight binding and the low energy (Dirac cone) approximations. Although only qualitative, this model nicely captures the main features observed on the STM data, and unambiguously shows that graphene's pseudospin is responsible for the suppression of the some quasiparticle interferences in monolayer graphene.  The model is in agreement with much more refined theory found in the literature, and a discussion is made on that point.
Finally, we present in section VI the experimental results obtained on BL graphene. The FT-LDOS map is not significantly affected by pseudospin effects, and qualitatively fits with the map discussed in section III. Once again, the quasiparticle dispersion relation for bilayer graphene is extracted from our data. 

\section{Experimental methods}

The epitaxial graphene samples were grown in ultra high vacuum (UHV) at NEEL Institute (France), using standard thermal decomposition of commercial SiC(0001) wafers (with 10$^{18}$ cm$^{-3}$ n-type doping), after a cleaning procedure at 900$^{\circ}$C under Si flux \cite{Forbeaux 1998, Mallet 2007, Varchon 2008}. The synthesis was optimized in order to get terraces of ML and BL graphene on the same surface, with typical width ranging between 50 to 250 nm. Both ML and BL terraces contain a significant amount of native atomic-sharp impurities (see sections III and V). The underlying C buffer layer was identified using low electron energy diffraction, showing a diagram with typical  SiC-($6\sqrt{3}\times6\sqrt{3})R30^{\circ}$ spots superimposed to the graphene 1$\times$1 spots  \cite{Owman 1996, Forbeaux 1998, Mallet 2007}.

The data presented here have been obtained on similar samples, using two different home made UHV microscopes working at 4-5K, one located at Max Planck Institute of Stuttgart (Germany), the other one at Universidad Autonoma de Madrid (Spain). After transport in atmosphere and transfer into the UHV systems, the samples were outgassed at 300-400$^{\circ}$C before cooling down. This procedure is sufficient for recovering high quality surfaces as checked by STM, with a defect density as low as the one found on the as-grown samples, thanks to the chemical inertness of the graphene layers. According to Tersoff and Hamman theory  \cite{TersoffHamann 1983}, the surface LDOS of our graphene samples was probed either in the constant current mode at low sample bias (a few mV), or at higher bias by performing  $dI/dV$ maps in open feedback-loop, using a lock-in technique, with frequency 2.3 kHz  and ac modulation of 2 mV applied to the sample. 2D fast Fourier transform (FFT) images, with square root normalization, were calculated from raw data STM images using the WSXM software  \cite{Horcas 2007}.

Particular care was taken on the data acquisition in order to achieve high resolution both in real and reciprocal space. One mandatory issue was to capture in the same STM image modulations with wavelength of few nanometers together with the atomic resolution. This is achievable by recording images of large areas (meaning terraces of width larger than 50 nm) with a high number of pixels, which is detrimental to the acquisition time (roughly 3 hours for a single $dI/dV$ map). We found that images of 100$\times$100 nm$^{2}$ with 4 megapixels were merely sufficient to evidence the fine structures on the 2D FFT maps discussed below. On such images, the $k$ resolution is intrinsically limited to 2$\pi$/100 $\simeq$ 0.063 nm$^{-1}$. For our low temperatures (4-5K), we use a 2 mV ac voltage modulation, which provides an energy resolution of the STM of $\sim$ 5 meV, reflected in an instrumental $k$ broadening similar to our size limitation, i.e. 
 0.06 nm$^{-1}$ (estimated from the dispersion relation of ML and BL graphene on SiC(0001) discussed in sections IV and VI). This value is 20 times smaller than the diameter of the ring-like pockets of the Fermi surface of ML and BL graphene on SiC(0001) (see section III). Another critical issue is the possible low frequency noise (from mechanical  or electronics sources) introduced by the experimental setup, which in most of reported works hinders the analysis of the center of the FFT images. As shown in section III, the instruments we used in the present study allow this kind of studies.

\section{Quasiparticle scattering in graphene: intra- and intervalley processes neglecting the impact of pseudospin}

We start this section by recalling briefly different properties of ideal ML graphene relevant for our study. The honeycomb structure of graphene is depicted on Fig. 1a, resulting from the superposition of two unequivalent triangular sublattices of $A$ and $B$ carbon atoms. Each type $B$ atom has 3 type $A$ nearest neighbors separated by vectors   $ \vec \tau _1 $, $\vec \tau _2 $ and  $\vec \tau _3 $. Figure 1b is a schematic view of the first Brillouin zone and of the Fermi surface (FS) for slightly doped graphene. At energies close to the Dirac point, the band structure of graphene consists of two Dirac cones at opposite $K $and $K'$ points of the Brillouin zone  \cite{Castro 2009}, with linear and isotropic dispersion $E(q_{E} )$ ($q_{E}$ is the modulus of the quasiparticle wavevector $ \vec q $ of energy $E$ measured from $K$ or $K'$ point, see inset of Fig. 1b). The FS is thus made of two points for neutral graphene, or two rings centered at $K$ and $K'$ points in the case of light doping ($K$ and $K'$ points are labeled $K _1$ and $K' _1$ on Fig. 1b, and the other points $K _2$ , $K'_2$, $K _3$, $K' _3$ are deduced from the symmetry of the reciprocal lattice). 

As already discussed in Ref. [\cite{Brihuega 2008}], ML and also BL graphene on SiC(0001) exhibit roughly the FS depicted on Fig. 1b, because of a n-type charge transfer from the buffer layer interface to the graphene layers. The band structure and the FS of ML and BL has been extensively studied by angle resolved photoemission (ARPES) groups  \cite{Zhou 2007, Bostwick 2007, Ohta 2006, Bostwick2 2007}: The radius $q _F$ of the FS pockets is close to 0.6 nm$^{-1}$, with a Dirac point at $\sim$  0.4 eV ($\sim $ 0.3 eV) below the Fermi energy $E_F$ for ML (BL) graphene. In the following, we will focus on the LDOS of ML and BL graphene on SiC(0001) close to $E_F$ in presence of a random distribution of impurities, and we shall concentrate on the corresponding FT-LDOS maps. As stated in the introduction, this problem has been addressed theoretically (in the case of a single impurity) by different groups  \cite{Bena 2008, PeregBarnea 2008, Bena 2009, Peres 2009, Bacsi 2010} and we shall refer to these works in section V. 

As a first step, we restrict ourselves to a qualitative picture which is successfully used for standard two dimensional electron gas (2DEG) such as noble metal and Be surfaces  \cite{Sprunger 1997, Petersen 2000} or high critical temperature superconductors  \cite{McElroy 2003}. It is derived from a simple analysis of the topology of the FS or of constant-energy contours for $E\ne E_F$. Static disorder in the 2DEG induces spatial modulations of the LDOS with wavevectors corresponding to vectors connecting different portions of the FS (or of the constant-energy contour).  From JDOS arguments, the scattering processes which have the largest weight in the LDOS are associated to good ÔnestingÕ vectors of the FS. In the simple case of the Shockley surface state of Cu(111), with a ring-like Fermi surface centered at $\Gamma$ point, backscattering processes (ie coupling between states $ \vec k _F $ and $- \vec k _F $) are the most significant: They contribute to an intensity ring of radius $2k_{F}$ ($2k_{F}$ ring in the following) at the center of the FT-LDOS map \cite{Petersen 1998, Petersen 2000}. This JDOS argument has been generalized to more complex FS \cite{McElroy 2003, Wang 2003, Vonau 2005}, for which the FS is no longer a simple contour centered on the $\Gamma$ point. However, it has been shown that this approach is insufficient in systems with large spin-orbit coupling, for which the wavefunction symmetry (in that case the electronic spin) hinders some scattering processes \cite{Petersen 1999, Pascual 2004}. 

Regarding graphene, if we neglect the possible impact of the wavefunction symmetries which shall be considered later, two different classes of elastic scattering processes are expected in ML and BL graphene, as depicted in Figs. 1c and 1d. On the one hand, long and short range scatterers generate intravalley scattering (coupling of states of a same FS pocket at $K_p$ or $K_p'$), with enhanced weight for backscattering processes (ie coupling between $ \vec q _F $ and  $- \vec q _F$ for all angles $\theta$) due to the circular shape of the pocket (Fig. 1c). Thus a $2q_{F}$ ring is likely to show up at the center of the FT-LDOS map (Fig. 1e). Also, replica of this ring are expected, centered at the first order spots of the reciprocal lattice, as shown on Fig. 1e. These rings result on the Bloch nature of the wavefunctions, as demonstrated in Refs. [\cite{Briner 1997, Petersen 2000}].

On the other hand, atomic sharp impurities also generate intervalley scattering processes (i.e. coupling between states of two different valleys at $K_p$ and $K_p'$ points). The latter processes yield the well documented ($\sqrt{3}\times\sqrt{3})R30^{\circ}$  superstructure on graphite/graphene STM images, as emphasized in the introduction  \cite{MizesFoster 1989, Ruffieux 2000, Ruffieux 2005, Fukuyama 2002, Kobayashi 2005, Sakai 2010, Mallet 2007, Rutter 2007, Brihuega 2008, Simon 2009, Dujardin 2010, Ye 2010, Ugeda 2010, Ugeda 2011, Martinez 2011}. Because of the topology of the FS pockets, coupling between states with opposite $ \vec q _F $  in each pocket (Fig. 1d) is highly favoured. For this example involving the pockets at $K _1$ and $K'_2$, the LDOS modulation shall have a wavevector 
$\overrightarrow {\Gamma {\rm K}}_{2} -2  \vec q _F$.
 Including all the orientations $\theta$ of $ \vec q _F $, a $2q_{F}$ ring centered at $\overrightarrow {\Gamma {\rm K}}_{2}$ is thus expected on the FT- LDOS map if the pseudospin is neglected, and  $2q_{F}$ rings should show up at other $K _p$, $K'_p$ points if we include the processes between all states of the FS (Fig. 1e).

 \section{High resolution STM results on monolayer graphene on SiC(0001) }
 
We want now to convince the reader that the FT-LDOS maps obtained by STM on ML graphene is markedly different from the schematic map sketched on Fig. 1e. We emphasize that a few publications have tried so far to give a description of the different features in the experimental FT-LDOS images of epitaxial graphene  \cite{Rutter 2007, Mallet 2007, Simon 2009, Dujardin 2010}. However, a poor resolution in $k$ space was partly limiting such analysis. In a report published in 2008, some of us have shown that it was possible to evidence fine structures in FT-LDOS data with improved quality  \cite{Brihuega 2008}. The data shown in the following have been obtained with an even better $k$ resolution (see method sections), which allows us to get one step further in our understanding of the quasiparticle scattering framework in graphene.

We first focus on the LDOS at $E \approx E_F$ of a 100$\times$100 nm$^2$ area of monolayer graphene. Following Tersoff and Hamann  \cite{TersoffHamann 1983}, this quantity can be obtained from a constant current STM image of the area at small sample bias, as shown in Fig. 2a. A clear triangular pattern with an almost SiC-6$\times$6 periodicity ($\sim$ 1.9 nm) can be appreciated in the image. However such pattern is not relevant for the present study, since it results from the interface contribution to the image  \cite{Chen 2005, RutterPRB 2007, Varchon 2008, Lauffer 2008}. Note that this image contains 2048$\times$2048 pixels, which is sufficient to resolve the graphene honeycomb atomic structure: Indeed, it shows up (together with the 6$\times$6 modulation) in numerical zooms taken at random spots on Fig. 2a, as shown on Fig. 2b. 

Figure 2c is the central part (40$\times$40 nm$^{-2}$) of the FT-LDOS calculated from Fig. 2a. For clarity, the dashed box drawn on the schematic FT map (obtained in section III, Fig. 1e) indicates the region of the reciprocal space which is probed.  Apart from the spots associated to the different orders of the 6$\times$6 superstructure, not relevant for the present study as explained above, two main features are evidenced on Fig. 2c: (i) There is no central ring of radius $2q_{F}$ (associated to intravalley backscattering) at the center of the diagram (see Fig. 2d, a magnified view of the boxed region in Fig. 2c) (ii) Rings of radius $2q_{F} \approx$ 1.1 nm$^{-1}$ , associated to intervalley scattering, are found centered at  $K _p$, $K'_p$  points, but with suppressed intensity along directions perpendicular to $\overrightarrow {\Gamma {\rm K}} _p$, $\overrightarrow {\Gamma {\rm K'}} _p$ vectors (Fig. 2e-g). These two features (i) and (ii) have been already reported in our previous letter \cite{Brihuega 2008} , but the data shown here have an improved resolution in $k$ space, in particular close to the  $K _p$, $K'_p$ points. The noise signal at the center of the FFT is also very weak, which allows us to rule out any possible central $2q_{F}$ ring. The features (i) and (ii) are in disagreement with the FT-LDOS map of Fig. 1e derived only from considerations on the shape of the FS, neglecting thus the role of the pseudospin. 

	One additional feature, not discussed in our previous report, shows up also on the FT-LDOS images:  $2q_{F}$ rings are present around the reciprocal lattice first-order spots. This is not shown on Fig. 2c since the size in $k$ space is too small, but such rings show up on Fig. 3a, obtained from measurements on a different ML terrace. Here, thanks to a terrace dimension close to 200 nm (which is among the largest terrace size reported so far for samples grown in UHV), we performed a 150$\times$150nm$^2$ constant current STM image (not shown) with 4096$\times$4096 pixels, at sample bias -10mV. The tip lateral resolution is excellent (see Fig. 3b, a 5$\times$5 nm$^2$ numerical zoom on the original 150$\times$150nm$^2$  STM image), and this allows us to analyse for the first time the intensity of the FT-LDOS at such high $k$ value. The area in k space shown on Fig. 3a is indicated by the dashed square drawn on the schematic FT. The center (0,0) of the FFT map is at the bottom of Fig. 3a. As on Fig. 2, we check the absence of any $2q_{F}$ ring at (0,0), together with the presence of anisotropic $2q_{F}$ rings at   $K _p$, $K'_p$ points. In addition, two of the first-order spots of the reciprocal lattice, labeled (1,0) and (0,1), show up on the figure. The two arrows point toward faint $2q_{F}$ rings centered at these points. These rings are expected on the FT map (Fig. 1e) derived in section II, as replica of the central ring associated to intravalley backscattering. As we will show in section V.C, the pseudospins cancels out the central ring, but not the replica rings at first-order spots of the reciprocal lattice. 
	
		From the different features found on the FT-LDOS maps shown on Figs. 2 and 3, we have exploited the $2q_{F}$ rings at $K _p$, $K'_p$ points to derive the low energy dispersion of the quasiparticles close to the Fermi energy. For that purpose, we made a series of $dI/dV$ maps of a ML terrace for voltages ranging between -125 to +125 mV, and processed them by FFT. Fig 4a shows the evolution of one of such anisotropic ring (indicated on the schematic FT map) with respect to the sample bias. We see that the anisotropy is robust, and that the radius $2q_{E}$ of the ring increases linearly with the voltage, and thus with energy $E$. The complete dispersion $E(q_{E})$, obtained from the rings around the 3 inequivalent $K_p$ points, is sketched on Fig. 4b. For voltages below -125 mV, we are not able to extract any reasonable $q_E$ value, because the associated wavelength is too large with respect to the image size. However, a linear fit gives a Dirac point located at -0.39 $\pm$ 0.01 eV, and a Fermi velocity $v_{F}$ =  (1.18 $\pm$ 0.04) $\times$ 10$^6$ m/s. The Fermi wavevector is $q_F$ = 0.53  $\pm$ 0.06 nm$^{-1}$. These values, obtained here with high accuracy, are very close to those derived from ARPES measurements \cite{Bostwick 2007, Zhou 2007} . In Ref. [\cite{Rutter 2007}] , a similar dispersion was obtained from STM measurements, but with a much larger $k$ uncertainty with respect to the present study. Note that in the $dI/dV(V)$ spectra measured at fixed tip position (Fig. 4c), a shallow minimum shows up at sample bias close $-0.4V$, i.e. at the Dirac energy derived in Fig. 4b. For ML graphene on SiC(0001), it is however difficult to extract properly the value of $E_D$ from such spectra,  due to the strong contribution of the interface states\cite{Rutter 2007, Lauffer 2008, Brar 2007} to the conductance signal at voltages $V<$-0.2 eV. 

In the next section of this manuscript, we will focus on the two major hallmarks found on the FT-LDOS maps in ML graphene: The absence of central $2q_{F}$ ring and the intensity anisotropy of the $2q_{F}$ rings at $K _p$, $K'_p$  points. As shown in section VI, these features are not observed for BL graphene, although the FS is roughly the same as for ML graphene. They are thus characteristic of the specific electronic properties of ML graphene.  As we already stated in Ref. [ \cite{Brihuega 2008}]  thanks to T matrix calculations  \cite{Bena 2008, Bena 2009}, the quasiparticle wavefunction symmetry (in other words the pseudospin), is the key ingredient for understanding such unique features in the FT-LDOS map. In the following, we introduce a simple model based on interferences between eigenstates of graphene obtained
in the tight binding approximation, which gives a simple demonstration of the impact of pseudospin on the quasiparticle interference framework. Our results will be discussed in the light of full theoretical predictions performed by other groups \cite{Bena 2008, PeregBarnea 2008, Bena 2009, Peres 2009, Bacsi 2010}.

\section{Discussion: Role of pseudospin on quantum Interferences in monolayer graphene}

We present in this section a simple and intuitive model to address the problem of single particle scattering off static impurities, and our purpose is to highlight the dramatic effect of the pseudospin in pristine graphene on the scattering mechanisms. 
In presence of defects, elastic scattering mixes eigenstates of the pristine system with the same energy, i.e. states that have different ${\vec k}$ wavevector located on the quasiparticle constant-energy contour \cite{Hoffman 2002, KodraAtkinson 2006, Simon 2007}. Thus, when computing the LDOS (which is proportional to the square modulus of the eigenstates of the disordered system)  in the vicinity of the impurities, one shall include terms of interference nature  
 $\psi _{\vec k}^* \left( {\vec r} \right)\psi _{\vec k'} \left( {\vec r} \right)$ (and its complex conjugated). Such terms correspond to scattering between arbitrary initial  $\psi _{\vec k} \left( {\vec r} \right)$ and final states $\psi _{\vec k'} \left( {\vec r} \right)$. A complete calculation of the LDOS should take into account the matrix elements which characterize the coupling for states ($\vec k$, $\vec k'$) as well as the boundary conditions at the defect sites, both being intimately linked to the nature, the symmetry, the strength of the impurities.

 For the sake of simplicity, and because the nature of the scatterers is usually unknown in real graphene systems, we shall focus in the following on the evaluation of the quantity  $\psi _{\vec k}^* \left( {\vec r} \right)\psi _{\vec k'} \left( {\vec r} \right)$ only. In that way, we are able to address the effect of the wavefunction symmetry on the interferences, without taking into account the specificity of the scatterer. Note that the details of the model are given in the supplemental material\cite{Supplemental Material}, and we give here only the main results. We refer to the basis and axis depicted on Figs. 1a and 1b.

\subsection{Wavefunctions in pristine graphene in the tight binding and the Dirac cone approximations}

The wavefunction in pristine graphene is written as a sum of two Bloch waves constructed on the two sublattices $A$ and $B$  \cite{BenaMontambaux 2009}:

\begin{equation}
\label{wavefunction}
\psi _{\vec k} \left( {\vec r} \right) = f_A \sum\limits_i {e^{i\mathord{\buildrel{\lower3pt\hbox{$\scriptscriptstyle\rightharpoonup$}} 
\over k} .\vec R_A^i } } \varphi \left( {\vec r - \vec R_A^i } \right) + f_B \sum\limits_i {e^{i\mathord{\buildrel{\lower3pt\hbox{$\scriptscriptstyle\rightharpoonup$}} 
\over k} .\vec R_B^i } } \varphi \left( {\vec r - \vec R_B^i } \right)
\end{equation}

$\varphi \left( {\vec r} \right)$  is the wavefunction of the $p_z$ orbital of each carbon atom. ${\vec R_A^i }$ (${\vec R_B^i }$)  is the position of atom $A$ ($B$) in the unit cell i. The complex quantities $f_A$ and $f_B$ coefficients are $\vec k$ dependent, which has been omitted here for simplicity. We use the simple notation  $\psi _{\vec k} \left( {\vec r} \right)=(f_A , f_B )$ in the following.

As in Ref. [ \cite{BenaMontambaux 2009}], we solve the Schr\"odinger equation in the tight binding approximation (with hopping between nearest neighbors only, and with on-site energies set to zero corresponding to the neutral graphene case). We also use the Dirac cone approximation \cite{Castro 2009, BenaMontambaux 2009} , by performing a low energy expansion to the first order in   $\left| {\vec q} \right|$ for a state in the pocket centered at $K _p$ (or  $K' _p$). We obtain the isotropic and linear dispersion relation of graphene:

\begin{equation}
\label{linear dispersion}
E(\mathord{\buildrel{\lower3pt\hbox{$\scriptscriptstyle\rightharpoonup$}} 
\over q} ) =  \pm \left| h \right| =  \pm t\frac{{ 3 }}{2}a\left| {\mathord{\buildrel{\lower3pt\hbox{$\scriptscriptstyle\rightharpoonup$}} 
\over q} } \right|
\end{equation}

with $\pm$ for electrons/holes ($a$=0.142 nm  and $t$=2.7 eV are respectively the distance and the hopping parameter between adjacent C atoms in graphene).

We also have a simple phase relation between $f_B$ and $f_A$ :

\begin{equation}
\label{phase relation}
f_B  =  \pm \frac{{h^* }}{{\left| h \right|}}f_A 
\end{equation}

with $\pm$ for electrons/holes.

The phase of $h$ is defined by (with $p$=1,2,3):

\begin{equation}
\label{phase of h}
h =  \mp \left| h \right|e^{ \pm i\theta } e^{ - i\frac{{2\pi }}{3}(p - 1)} 
\end{equation}
with -/+ for states of a pocket at $K _p$/$K'_p$ point \cite{BenaMontambaux 2009}.

Equation (3) implies that $f_B$ and $f_A$ are equal in modulus, and thus that the LDOS is the same on $A$ and $B$ sublattices. Moreover, equations (3-4) show that the phase relation between $f_B$ and $f_A$ depends in a peculiar way on the orientation of wavevector $\vec q$ (i.e. on the quasiparticle momentum), due to the $e^{ \pm i\theta }$ term in equation (4). This phase relation is depicted in the literature as a pseudospin, whose orientation is either parallel or antiparallel to the momentum, defining an electronic chirality \cite{Katsnelson 2006, Castro 2009}. The pseudospin texture in ML graphene is depicted on Fig. 5a. For a given state of the $K$ valley at energy above $E_D$, the pseudospin is aligned to the wavevector $\vec q$. Importantly, the pseudospin associated to opposite wavevector $-\vec q$ in the same valley is reversed (Fig. 5b): The phase shift between $f_B$ and $f_A$ changes its sign when $\theta$ is changed into $\theta + \pi$ in expressions (3) and (4). This implies that in presence of long range disorder (conserving the pseudospin), intravalley backscattering is not possible  \cite{Ando 1998, Katsnelson 2006}. Equations (3) and (4) also show that the orientation of the pseudospin is reversed for energies below the Dirac point, as schematized in Fig. 5b. Moreover, the pseudospin  texture is reversed between the two valleys (Fig. 5a). 
In the following, we will use the pseudospin term to refer to this peculiar symmetry property of quasiparticles wavefunctions in graphene, which is directly associated to the honeycomb structure.

\subsection{Expression of the interference term $\psi _{\vec k}^* \left( {\vec r} \right)\psi _{\vec k'} \left( {\vec r} \right)$}

We consider scattering processes between wavefunctions $\psi _{\vec k} \left( {\vec r} \right) = (f_A , f_B )$
and $\psi _{\vec k'} \left( {\vec r} \right) = (f'_A , f'_B )$, defined in section V.A, and we calculate the interference term $\psi _{\vec k}^* \left( {\vec r} \right)\psi _{\vec k'} \left( {\vec r} \right)$ (the wavevectors $\vec k$ and $\vec k'$ lie on a constant-energy contour of energy $E$).
We obtain the following expression (see Supplemental Material\cite{Supplemental Material}):

\begin{equation}
\label{Interference term}
\psi _{\vec k}^* \left( {\vec r} \right)\psi _{\vec k'} \left( {\vec r} \right) = \frac{1}{{2N}}e^{i\varphi } e^{i(\vec k' - \vec k).\vec r} \sum\limits_{\vec G} {\tilde F_{\vec G}^{\vec k' - \vec k} } e^{i\vec G.\vec r} \left( {1 + \frac{{hh'^* }}{{E^2 }}e^{i\vec G.\vec \tau _1 } } \right)
\end{equation}

where the sum runs over all wavevectors $\vec G$ of the reciprocal lattice. ${\tilde F_{\vec G}^{\vec k' - \vec k} }$ is the $\vec G$th Fourier component of a function $F_A (\vec r)$ defined on sublattice $A$ (see Supplemental Material\cite{Supplemental Material}). The angle $\varphi$   is defined by $f_A^* f'_A  = \left| {f_A^* } \right|\left| {f'_A} \right|e^{i\varphi } $. 2N is the total number of atoms in the system.

Interestingly, equation (5) shows that the interference term $\psi _{\vec k}^* \left( {\vec r} \right)\psi _{\vec k'} \left( {\vec r} \right)$ can be written as a sum of plane waves $e^{i(\vec k' - \vec k + \vec G).\vec r} $. Consequently its Fourier transform (and hence the FT of the LDOS) should be peaked at wavevectors $\vec k' - \vec k + \vec  G$, with an intensity modulated by the prefactor terms in (5). The most relevant is the term in bracket in equation (5), deemed Ôintensity factorÕ in the following. In table 1, we evaluate this quantity for different initial and final states ($\vec k$ ,$\vec k'$), and different $\vec G$ vectors. Since we want to refer to real experiments, which are limited in $k$ space, we retain $\vec G = \vec 0$ and $\vec G$ vectors with modulus  $\left| {\vec G} \right| = \left| {\vec a^* } \right|$. In the following, we discuss these results separating intravalley and intervalley processes.

\subsection{Intravalley backscattering contribution}

We choose initial and final states in a same valley at $K_p$ (the results are identical for the valley at  $K'_p$), and we consider the most relevant processes, i.e. backscattering processes ($\theta' = \theta  + \pi$). Hence $\vec k' - \vec k=\vec q' - \vec q =  -2 \vec q$. We first evaluate the low frequency component in (5) at $\vec G = \vec 0$, which should give intensity in the FT-LDOS map at  $-2 \vec q$, at the vicinity of the center $\Gamma$. From table 1, we see that the intensity factor is strictly zero, whatever the direction of $\vec q$. Consequently, there will be no circle of radius $2q_E$ at the center of the FT-LDOS map. This is a consequence of the pseudospin, i.e. the symmetry of the quasiparticle wavefunctions given in equations (3-4), which leads to the cancellation of the intensity factor in equation (5) for $\vec G = \vec 0$ (see Supplemental Material\cite{Supplemental Material}). The quasiparticle interferences at wavector  $2q_E$ associated to the intravalley backscattering processes are thus annihilated by this intrinsic property of graphene, independently of the nature of the scatterer. This is what we observe experimentally on Figs. 2 and 3.

Because of the phase term $e^{i\vec G.\vec \tau _1 }$ in equation (5), the replica of the intravalley backscattering term at  $\vec G \ne \vec 0$ do not vanish. As sketched in Table 1, we find for the first-order components $\vec G = \pm$  $\vec a^*$, $\vec G = \pm$  $\vec b^*$,  $\vec G =  \pm  \left( {\vec a^*  - \vec b^* } \right)$ that the intensity factor is a non zero constant for backscattering at any angle $\theta$. Consequently a replica signal in the FT- LDOS map is expected, showing up as $2q_E$ rings around each first order spots of the reciprocal lattice. This is precisely what we obtain in our highly-resolved experimental data (Fig. 3).

At that point, it is necessary to make a connection with the recent theoretical calculations mentioned in the introduction (section I). Based on more elaborated models using Greens function formalism, FT-LDOS maps of graphene in presence of a single impurity have been calculated by several groups  \cite{Bena 2008, PeregBarnea 2008, Bena 2009, Peres 2009}. The presence of  $2q_E$ rings at lattice spots and the lack of central  $2q_E$ ring, which we experimentally observe,  are also predicted in these calculations. Importantly, the suppression of the central  $2q_E$ ring (and hence of the interferences with wavector  $2q_E$) related to intravalley backscattering, exist only if both the $A$ and $B$ sublattices are taken into account in the calculation of the FT-LDOS maps  \cite{Bena 2008, PeregBarnea 2008, Bena 2009, Peres 2009}. As highlighted by the authors, the interferences with  $2q_E$ wavevectors due to a delta impurity exist on each sublattice but are shifted by $\pi$ from one to the other, and thus the two contributions cancel each other when the two lattices are taken into account \cite{Note A}. It is straightforward to check that we get the same result with our model: If we evaluate separately the zeroth $\vec G$ order component of $\psi _{\vec k}^* \left( {\vec r} \right)\psi _{\vec k'} \left( {\vec r} \right)$ on $A$ and $B$ sublattices, we find that the two quantities are opposite, and thus cancel each other once they are summed.  

Although oversimplified, our model nicely explains why the pseudospin induces the cancellation of the  $2q_E$ oscillations when both sublattices are considered, and this irrespective of the nature of the scatterer. We agree with the authors of Refs. \cite{Bena 2008, PeregBarnea 2008, Bena 2009, Peres 2009, Bacsi 2010} that in STM measurements, the  $2q_E$ oscillations are possibly present with opposite phase on $A$ and $B$ sublattices, although it appears very difficult to measure. As explained below, this is not due to any experimental limitation, since we performed the measurements with sharp tips enabling to clearly evidence the graphene honeycomb lattice, combined with excellent energy and wavevector resolution.

In order to understand what is possibly measurable by STM, it is worth to illustrate the cancellation of the $2q_E$ oscillations on the basis of artificially generated LDOS images obtained by simple combination of cosine functions. Such functions are used to mimic the two sublattices and the possible LDOS oscillations due to one single impurity located at the center of the images (we are obviously not considering here the $\sqrt{3}\times\sqrt{3} R30^{\circ}$ oscillations).
On Fig. 6a, we focus on the situation where sublattice $A$ only is involved\cite{Note B}: The triangular lattice (obtained by a product of 3 cosines) is multiplied by a radial cosine function with wavevector  $2q_F$ = 1.2 nm$^{-1}$. The $2q_F$ radial oscillation shows up on the 100$\times$100 nm$^2$ image (Fig. 6a), and a numerical zoom performed on the 8$\times$8 nm$^2$ boxed area reveals the $A$ sublattice (Fig. 6b). A profile performed along a row of A atoms is shown on Fig. 6c, highlighting the $2q_F$ modulation. The 2D-FFT of Fig. 6a is shown on Fig. 6d. It contains the features discussed in section II: a central $2q_F$ ring, and $2q_F$ rings around first order lattice spots \cite{Note C}. 

We focus now on Fig. 6e, where both sublattices are included. Radial oscillations with wavectors $2q_F$ are introduced on each sublattice, but with opposite phase \cite{Note B}. As a result, the $2q_F$ oscillation is completely smeared out on the real space image, and no central $2q_F$ ring is found on the 2D FFT (Fig. 6h). This implies that for perfect ML graphene, STM will never observe any signal coming from intravalley scattering processes in the central region of the 2D FFT, independently of the microscope resolution. Fig. 6f is a zoom on the dashed square region of Fig. 6e. Although no $2q_F$ oscillation shows up on the image, such oscillations are revealed on the profiles taken along rows of A or B atoms (Fig. 6g), with the introduced $\pi$ phase shift between them. 

Interestingly, Fig. 6f  shows that the graphene honeycomb pattern is not perfectly uniform: Indeed, as we can also deduce from the work of Peres et al. \cite{Peres 2009}, patches with almost honeycomb contrast alternate with areas showing a faint  $AB$ asymmetry, with a period which is twice the wavelength of the $2q_E$ oscillations. In principle, STM should be able to detect such regions with $AB$ asymmetry, providing that the asymmetry is significantly large. From Ref. [\cite{Peres 2009}], this asymmetry is in fact very weak away from the impurity, compared to the asymmetry found on bilayer graphene with Bernal stacking \cite{Wang 2007}. In our STM data on ML graphene (Figs 2 and 3), we have no indication of such an atomic contrast. In principle, another way to extract the $2q_F$ oscillations of one sublattice would be to do profile measurements along one single $A$ (or $B$) atomic row as on Fig. 6g. However, in the present case, even on our best images, we get no significant result from such profiles, the measurement being complicated by the SiC-6$\times$6 modulation due to the interface.

\subsection{Intervalley scattering contribution}

We now consider the intervalley scattering processes, which couple states of two neighboring pockets, for instance $K_1$ and $K'_2$. As stated in section II, the most relevant processes imply states with opposite $\vec q$ vectors, hence $\vec k' - \vec k = \overrightarrow {\Gamma {\rm K}} _2  -2 \vec q$. From equation (5), such processes will give signal intensity in the FT-LDOS map close to   $\overrightarrow {\Gamma {\rm K}} _2 +  \vec G$ for all $\vec G$ values of the reciprocal lattice. We restrict ourselves to the three terms $\vec G = \vec 0$, $\vec G = - \vec b^*$, $\vec G = \vec a^* - \vec b^*$, which give signal intensity in the first Brillouin zone, around $\overrightarrow {\Gamma {\rm K}} _2$, $\overrightarrow {\Gamma {\rm K}} _3$ and $\overrightarrow {\Gamma {\rm K}} _1$ respectively. From table 1, we see that the intensity factor is generally non-zero and depends on the orientation $ \theta$ of $\vec q$ vector. It follows that in the FT-LDOS map,  $2q_E$ rings with anisotropic intensity are expected around points $K_1$, $K_2$, $K_3$. Most importantly, the intensity factor has zeroes at specific angles $ \theta_0$ and  $ \theta_0 +  \pi$ (for instance at   $ \theta_0 =  \pi/6$ and $ \theta_0 +  \pi =  -5\pi/6$ for $\vec G = \vec 0$), which implies that the $2q_E$ rings are split in two parts. As detailed in the Supplemental Material\cite{Supplemental Material}, the intensity the  $2q_E$ ring centered at $K_p$  is suppressed in the direction perpendicular to ($\overrightarrow {\Gamma {\rm K}} _p$).

More generally, equation (5) demonstrates that intervalley scattering processes between states with opposite  $\vec q$ vectors  contribute to $2q_E$ rings around the $K_p$ and $K'_p$ points in the FT-LDOS map. The intensity of such rings is suppressed in the directions perpendicular to ($\overrightarrow {\Gamma {\rm K}} _p$) or ($\overrightarrow {\Gamma {\rm K}} _p'$). This suppression is once again due to the wavefunction symmetry (pseudospin), which gives the prefactor term in equation (5). The ring anisotropy around the  $K_p$ and $K'_p$ points is clearly revealed in our STM measurements (section III and Ref.  [\cite{Brihuega 2008}]). It is also predicted in Refs.  \cite{Bena 2008, PeregBarnea 2008, Brihuega 2008, Bena 2009, Peres 2009}, although the link with the pseudospin is not as straigthforward as in the present work.

\subsection{Concluding remarks about the model }

With our calculations, we demonstrate the impact of the graphene$'$s pseudospin on the quasiparticle interferences, and show how it affects the FT-LDOS maps. The main results of the model are summarized on Fig. 5c, which is the schematic FFT map derived from JDOS consideration in section II, corrected by the pseudospin effects described here. 
The agreement between Fig. 5c and the experimental FT-LDOS maps shown in Fig. 2 and 3 is only qualitative, but the model nicely captures all the main features observed on our data. The theoretical predictions are valid for all kind of scatterers, which is satisfying since the real nature of the impurities is usually unknown as in the present study. It is also interesting to do the calculation in the case of an asymmetric monolayer graphene (see Supplemental Material\cite{Supplemental Material}): we find that the vanishing intensities (the central  $2q_E$ ring and the nodes of  $2q_E$ rings at $K$,$K'$ points) are restored with increasing the difference between onsite energies of $A$ and $B$ sites.

\section{STM results and discussion for bilayer graphene}

This section is devoted to a brief description of quasiparticle interferences in BL graphene (in the case of Bernal stacking).
 As for ML graphene, low-energy quasiparticles in BL graphene  also present a pseudospin degree of freedom, associated with the complex wavefunction amplitudes on the two layers. The pseudospin is
 linked to the momentum in a different way than in ML graphene  \cite{McCannFalko 2006, Katsnelson 2007}. This is illustrated on Fig.7, where the pseudospin textures for ML (Fig. 7a) and BL graphene (Fig. 7b) are shown (the case of standard 2D electron gas, without pseudospin, is also shown on Fig. 7c). As depicted on Fig. 7b, the pseudospins of states of opposite $\vec q$ vectors are parallel, and thus the intravalley backscattering processes are promoted \cite{Katsnelson 2007}, as in standard 2D electron gas where no pseudospin is present. This should be reflected in the QIs pattern probed by STM, as discussed in section III.

Hence, we focus now on the experimental data obtained on BL graphene terraces on SiC(0001). To obtain the images shown on Fig. 8, a 50$\times$50 nm$^2$ constant current image (not shown) was recorded at sample bias -25 mV. A mathematical 5$\times$5 nm$^2$ zoom of such image is shown on Fig. 8a, displaying a triangular pattern characteristic of bilayer graphene with Bernal stacking \cite{Mallet 2007, Rutter 2007, Brar 2007, Lauffer 2008, RutterJVST 2008}.  Fig. 8b is a 50$\times$50 nm$^2$ $dI/dV$ image, recorded simultaneously with the topography, corresponding to a LDOS map at -25 meV below $E_F$. The image is clearly dominated by long range oscillations with period of few nm, associated to intravalley backscattering processes  \cite{Rutter 2007, Brihuega 2008}.

 This is confirmed by the 2D FFT of Fig. 8b shown on Fig. 8c. 
Contrary to the monolayer case (see the 2D FFT map displayed on Fig. 2c for comparison), a  $2q_E$ ring associated to intravalley backscattering is found at the center of the image, and complete  $2q_E$ rings associated to intervalley scattering are present at $K$,$K'$ points. These results confirm the measurements already reported in Ref.  \cite{Brihuega 2008}, and are in good agreements with T matrix calculations  \cite{Brihuega 2008, Bena 2008}. As mentioned at the beginning of this section,  the pseudospin in bilayer graphene does not hinder the backscattering processes \cite{Katsnelson 2007}, and thus the central  $2q_E$ ring is expected on the FT-LDOS map. Using similar calculations as those in section IV, it is possible to check that the pseudospin in BL graphene restores the central  $2q_E$ ring on the FT-LDOS map, and also the intensity isotropy of  the $2q_E$ rings at $K$,$K'$ points.

	On Fig. 9a, we present on the top row a series of $dI/dV$ images of the same terrace taken at different sample biases ranging from -250 to +50 mV. 2D FFT maps of these images have been calculated, and we have extracted for each image two zoom-in pictures: One is a  $2q_E$ ring at $K$ point (Fig. 9a, middle row) and the other is the central $2q_E$ ring (Fig. 9a, bottom row),
as indicated by the left side schematics. We find a concomitant increase of the rings radius with the voltage (energy), and we can extract the dispersion relation for BL graphene on SiC(0001) shown in Fig. 9b. Our data are consistent with the theoretical low-energy dispersion of bilayer graphene \cite{McCannFalko 2006, Wang 2007}, taking into account a n-type doping from the interface ($E_D$ =-0.3 eV), and a 0.1 eV bandgap at $E_D$ due to a different doping of the two layers  \cite{Ohta 2006, Ohta 2007}.  The best fit to $E(q_E)$ data of Fig. 9b (plain curve) is obtained with a Fermi velocity $v_F$ = 1.21 10$^6$ m/s  and an interlayer hopping parameter $\gamma_1$ = 0.38 eV, close to the values derived from ARPES measurements \cite{Ohta 2007}. 
	The energy bandgap of $\sim$ 0.1 eV shows up in $dI/dV$ spectra as a dip around -0.3eV, as shown on Fig. 9c. Note that our spectrum is very similar to the spectra reported for BL graphene on SiC(0001) in Refs. \cite{Brar 2007, Lauffer 2008}, including the conductance dip around zero bias whose origin is still debated. The shift of the Dirac point with respect to  ML graphene  (see Fig. 4b and 4c) is consistent with ARPES data and results from charge transfer and screening effects as discussed in Ref.  [ \cite{Ohta 2007}].

\section{Conclusion}	

We have performed a complete  study of the quasiparticle interferences (QIs)  in epitaxial graphene on SiC(0001), by using low temperature scanning tunneling microscopy and spectroscopy. This technique is carried out to map the spatial modulations of the local density of states (LDOS) of monolayer (ML) and bilayer (BL) graphene in presence of native disorder. The high resolution achieved here allows a thorough analysis of the different ring-like features found in the two-dimensional Fourier transform of the data. We introduce a simple model which nicely captures the main features observed on the FT LDOS map, and which unambiguously demonstrate the impact of the pseudospin degree of freedom on the QIs pattern. We also derive with a great accuracy the quasiparticle dispersion relation for both ML and BL graphene on SiC(0001) in the vicinity of the Fermi level. 

 Our main results  are summarized on Fig. 10, which reproduces some of the figures described in the manuscript, in a fashion which favours a quick comparison between the ML and BL graphene systems.

\section{Acknowledgements}

P.Mallet, I. Brihuega, M.M. Ugeda, J.M. Gomez Rodriguez and J.Y. Veuillen acknowledge the PHC Picasso program for financial support (project n¡ 22885NH). Partial financial support from MICINN (Grants No. MAT2010-14902 and CSD2010-00024) and from CAM (S2009/MAT-1467) is also gratefully acknowledged.

\section{Table}

\begin{tabular}{|c|c|c|}

\hline
\hline
  & Spot location in the FT-LDOS map & Intensity factor \\ 
   & $\vec k' - \vec k + \vec  G$  & $ {1 + \frac{{hh'^* }}{{E^2 }}e^{i\vec G.\vec \tau _1 }}  $  \\
\hline
\hline
\multicolumn{3}{|c|}{ Intravalley backscattering (pocket at $K_p$):  $\vec k$=$\overrightarrow {\Gamma {\rm K}} _p +  \vec q $, $\vec k'$=$\overrightarrow {\Gamma {\rm K}} _p -  \vec q $}  \\
\hline
 $\vec G = \vec 0$ &  $ -2 \vec q$ & 0  \\
\hline
$\vec G = \pm \vec a^*$  &  $\pm \vec a^* -2 \vec q$ &  $1 - e^{ \pm i\frac{{2\pi }}{3}}$  \\
\hline
$\vec G = \pm \vec b^*$  &  $\pm \vec b^* -2 \vec q$ &  $1 - e^{ \pm i\frac{{4\pi }}{3}}$  \\
\hline
$\vec G =  \pm \left( {\vec a^*  - \vec b^* } \right)$  &  $\pm \left( {\vec a^*  - \vec b^* } \right) -2 \vec q$ &  $1 - e^{ \mp i\frac{{2\pi }}{3}} $  \\
\hline
\hline
\multicolumn{3}{|c|}{  Intervalley backscattering between states of pockets $K_1$ and $K'_2$: $\vec k$=$\overrightarrow {\Gamma {\rm K}} _1 +  \vec q $, $\vec k'$=$\overrightarrow {\Gamma {\rm K'}} _2 -  \vec q $}  \\
\hline
$\vec G = \vec 0$ & $\overrightarrow {\Gamma {\rm K}} _2  -2 \vec q$ & $\sim \sin \left( {\theta  - \frac{\pi }{6}} \right)$  \\
\hline
$\vec G = - \vec b^*$ & $\overrightarrow {\Gamma {\rm K}} _3  -2 \vec q$  & $\sim \sin \left( {\theta  - \frac{5\pi }{6}} \right)$\\
\hline
$\vec G =  \vec a^* - \vec b^*$ & $\overrightarrow {\Gamma {\rm K}} _1  -2 \vec q$  & $\sim \sin \left( {\theta  - \frac{\pi }{2}} \right)$\\
\hline
\hline

\end{tabular}

Table1: Summary of the results obtained with our model. We calculate the $\vec G$th component of the interference term $\psi _{\vec k}^* \left( {\vec r} \right)\psi _{\vec k'} \left( {\vec r} \right)$, for different $\vec k$, $\vec k'$ states. This quantity corresponds to a spot intensity in the FT-LDOS map at $\vec k' - \vec k + \vec G$ (Ôspot locationÕ colomn). This intensity is modulated by the so-called 'intensity factor' (last column) defined in the main text.

\section{Figure captions}

Figure 1 (Color online): (a) The honeycomb structure of monolayer (ML) graphene. The unit cell includes one carbon atom on each $A$ and $B$ sublattice (two C atoms per unit cell). (b) The first Brillouin zone and the Fermi surface of n-doped ML graphene. Inset: The low energy band structure of ML graphene (Dirac cone), with a linear and isotropic dispersion at $K$ or $K'$  points. $E_D$ and $E_F$ are respectively the Dirac and Fermi energies. (c) and (d) Schematic of the elastic scattering events in doped ML/BL graphene, divided into intravalley (c) and intervalley (d) processes. The arrows correspond to the wavevectors of the associated LDOS modulations. (e) Schematic of the expected two dimensional Fourier transform map of the LDOS, including all the processes depicted in (c) and (d), and neglecting the pseudospin (see the main text). 

Figure 2  (Color online) (a) A 100$\times$100 nm$^2$ constant current STM image on ML graphene on SiC(0001). Sample bias: -4 mV. Number of pixels: 2048$\times$2048 (b) A 5$\times$5 nm$^2$ numerical zoom of (a), showing a hexagonal atomic pattern of period ~0.24 nm characteristics of the graphene's honeycomb structure. A long range periodic superstructure (period 1.9 nm) is also present, inferred to the interface with the buffer layer (see text). (c) Two-dimensional fast Fourier transform (2D FFT) of (a). Image size: 40 $\times$ 40 nm$^{ - 2}$. (d) Central area of (c) showing the absence of intensity ring with radius $2q_F$. (e-g) Zoom-in on the three $2q_F$ outer rings in (c) indicated by arrows. Image sizes are 5 $\times$ 5 nm$^{- 2} $ for (d-g). 

Figure 3.  (Color online) (a) Detail of the 2D FFT of a 150 $\times$ 150  nm$^2$ constant current STM image obtained at sample bias -10mV on ML graphene on SiC(0001).  Image size 40 $\times$ 40 nm$^{ - 2}$. The center $k$=0 of the FFT is labeled (0,0), and first order spots of the reciprocal lattice are labeled (1,0) and (0,1). The arrows point towards faint rings of radius $2q_F$ centered at these two points. (b) a 5$\times$5 nm$^2$ numerical zoom of the 150 $\times$ 150 nm$^{  2} $ image.

Figure 4.  (Color online) (a) Sample-bias dependence of a $2q_E$ ring at $K$ point in the FT-LDOS maps, obtained from a series of 50$\times$50 nm$^2$ $dI/dV$ images (not shown). Each image of (a) has a size 5 $\times$ 5 nm$^{ - 2}$. (b) Dispersion relation $E(q_E)$ extracted from the radial average of the rings shown in (a). A linear fit is displayed in plain lines, yielding to an estimation of the Fermi velocity $v_F$ and of the Dirac energy $E_D$. (c) A typical $dI/dV(V)$ spectrum obtained at fixed tip position, with open feedback loop. The arrow points towards a shallow minimum of the conductance curve at sample bias corresponding to the Dirac energy $E_D$ derived in (b). Stabilisation parameters: Sample bias:  + 350 mV, Tunneling current: 0.15 nA.

Figure 5.  (Color online) (a) Schematic Fermi surface of n-doped ML graphene, featuring the pseudospin orientations (red arrows) in the two unequivalent valleys at $K$ and $K'$ points. (b) Schematic of the Dirac cone at K point showing the pseudospin orientations (red arrows) for states at energies above and below the Dirac point. (c) Expected FT-LDOS map taking into account the FS topology and the pseudospin (see text).

Figure 6. (Color online) Illustration of the QIs associated to intravalley scattering off a single atomically-sharp impurity. (a) Simulation of the $2q_F$ LDOS oscillation generated on one sublattice only (sublattice $A$), with the impurity located at the center of the image on a site $A$. Image size 100$\times$100 nm$^2$, 2048$\times$2048 pixels. (b) Numerical zoom of the area limited by the dashed box on (a). (c) Lateral profile along a row of A atoms performed on (b). (d) 2D FFT of (a). (e) same representation as (a), considering both $A$ and $B$ sublattices, with $2q_F$ oscillations shifted by $\pi$ between the two sublattices. No long range oscillation show up on this image, in contrast with (a). (f)  Numerical zoom of the area limited by the dashed box on (e). (g) Lateral profiles performed on (f) along a row of A atoms and along a row of B atoms. (h) 2D-FFT of (e). See Ref. [\cite{Note B}] for the exact quantity mapped on (a) and (e).

Figure 7. (Color online) (a-b) Schematic representations of the low energy band structure and of the pseudospin texture (red arrows) in (a) Monolayer graphene (b) Bilayer graphene. (c) Band structure of a standard 2D electron gas. Note that in this latter case, the band is centered on the $\Gamma$ point of the Brillouin zone.

Figure 8.  (Color online) (a) A 5 $\times$ 5 nm$^2$ numerical zoom of a 50 $\times$ 50 nm$^2$ constant current STM image (not shown) recorded on a BL terrace, at sample bias -25 mV. The atomic triangular pattern on (a) is the hallmark for bilayer graphene with Bernal stacking. (b) A 50$\times$50 nm$^2$ $dI/dV$ map of the BL terrace at sample bias -25mV. (c) 40$\times$40 nm$^{ - 2}$ 2D FFT image calculated from (b). Both intravalley and intervalley $2q_E$ rings show up, respectively at the center and at $K$ ($K'$) points of the diagram. 

Figure 9.  (Color online) (a) Top row: Series of 50$\times$50 nm$^2$ $dI/dV$ maps of a BL terrace at different sample biases. Middle  and bottom rows: corresponding 2D FFT maps, magnified respectively on a $2q_E$ ring at $K$ point and on the central $2q_E$ ring . Images size: 5 $\times$ 5 nm$^{ - 2}$. (b) Dispersion relation extracted from the average radius of the rings displayed on (a). The plain curve is a fit of the data for a n-doped asymmetric bilayer, with parameters given in the table, close to the parameters derived from ARPES measurements \cite{Ohta 2007} (c) A $dI/dV(V)$ spectrum obtained at fixed tip position, with open feedback loop. A dip of width 0.1eV centered at -0.3 eV reflects the energy bandgap induced by the doping asymmetry between the two graphene layers. Stabilisation parameters: sample bias:  + 350 mV, Tunneling current: 0.15 nA.

Figure 10.  (Color online) Summary table of the main results obtained in this study. (a) Schematic pseudospin texture of ML. (b) Schematic FT LDOS map taking into account the Fermi surface topology and the pseudospin of ML, derived from our model. (c) FT LDOS map obtained from STM measurements on ML graphene on SiC(0001) at 5K. Note the good agreement between (b) and (c). (d) Dispersion relation derived from the STM data on ML.
(e-h) same as (a-d) but for BL graphene. Note the correspondance between (f) and (g): the pseudospin has no significant impact for the BL, contrary to the ML case.

\newpage
\begin{figure}[!h]
\begin{center}
\includegraphics[width=18 cm,clip]{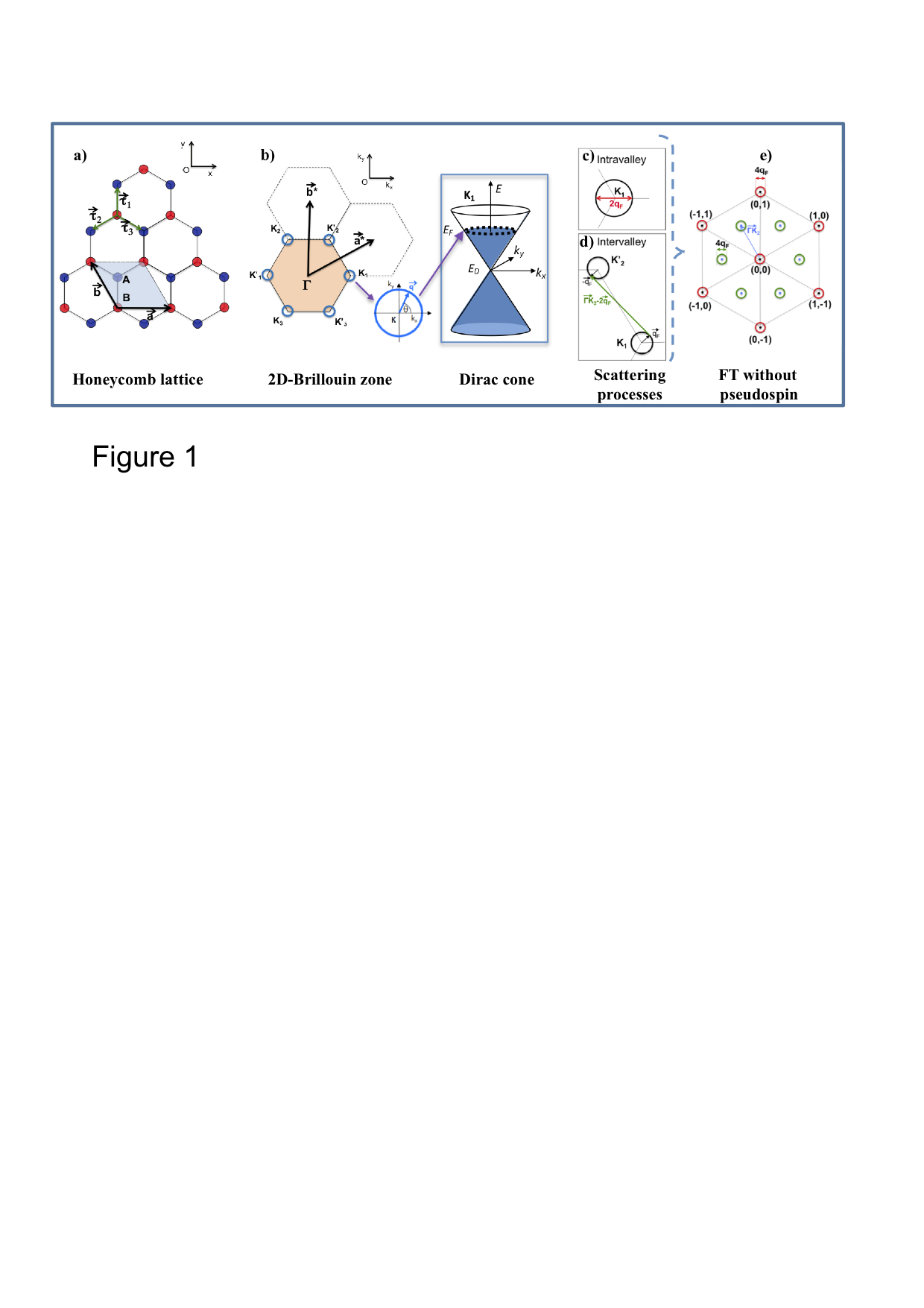}
\end{center}
\label{Figure 1}
\end{figure}

\newpage
\begin{figure}[!h]
\begin{center}
\includegraphics[width=18 cm,clip]{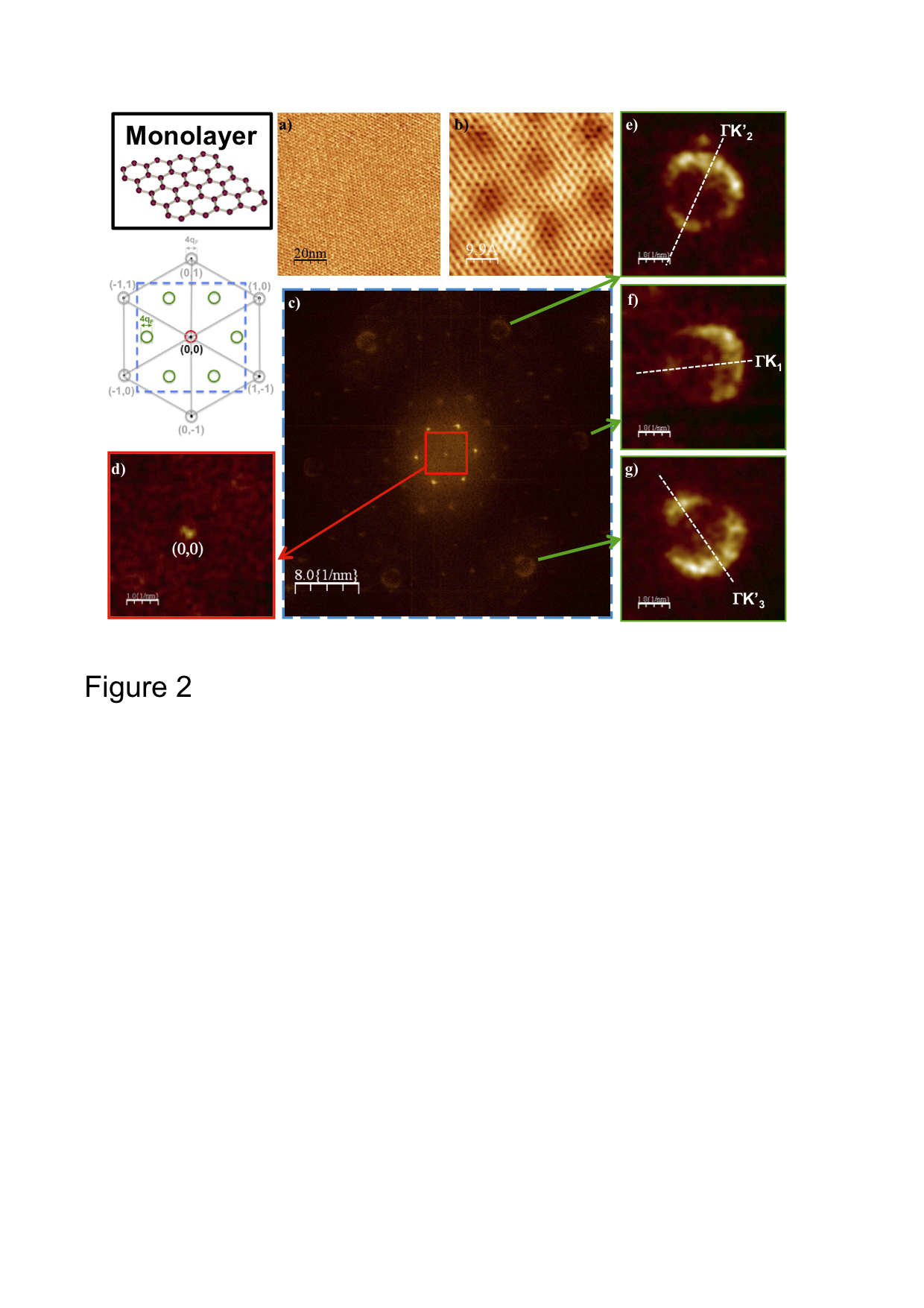}
\end{center}
\label{Figure 1}
\end{figure}

\newpage
\begin{figure}[!h]
\begin{center}
\includegraphics[width=18 cm,clip]{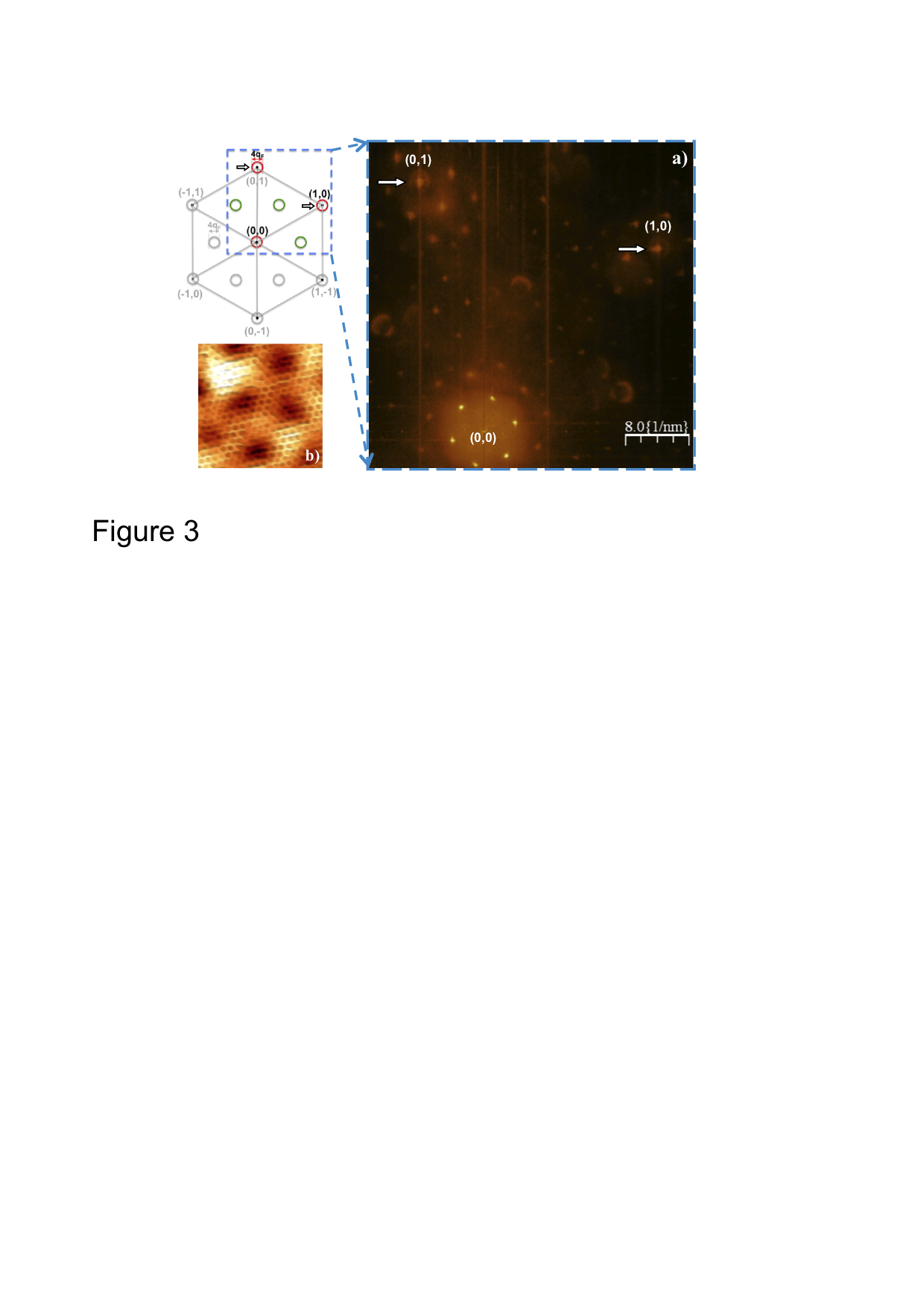}
\end{center}
\label{Figure 1}
\end{figure}

\newpage
\begin{figure}[!h]
\begin{center}
\includegraphics[width=18 cm,clip]{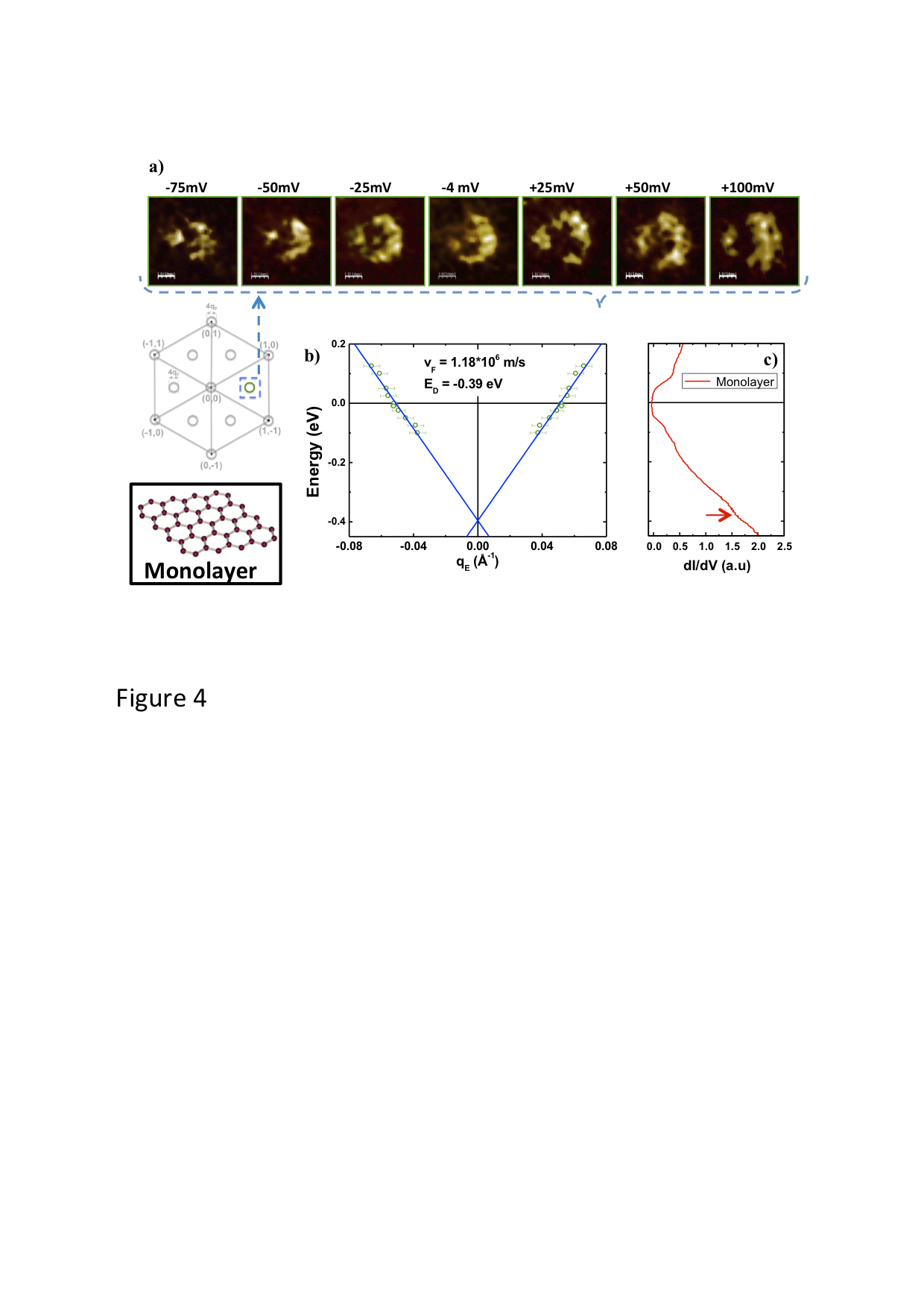}
\end{center}
\label{Figure 1}
\end{figure}

\newpage
\begin{figure}[!h]
\begin{center}
\includegraphics[width=18 cm,clip]{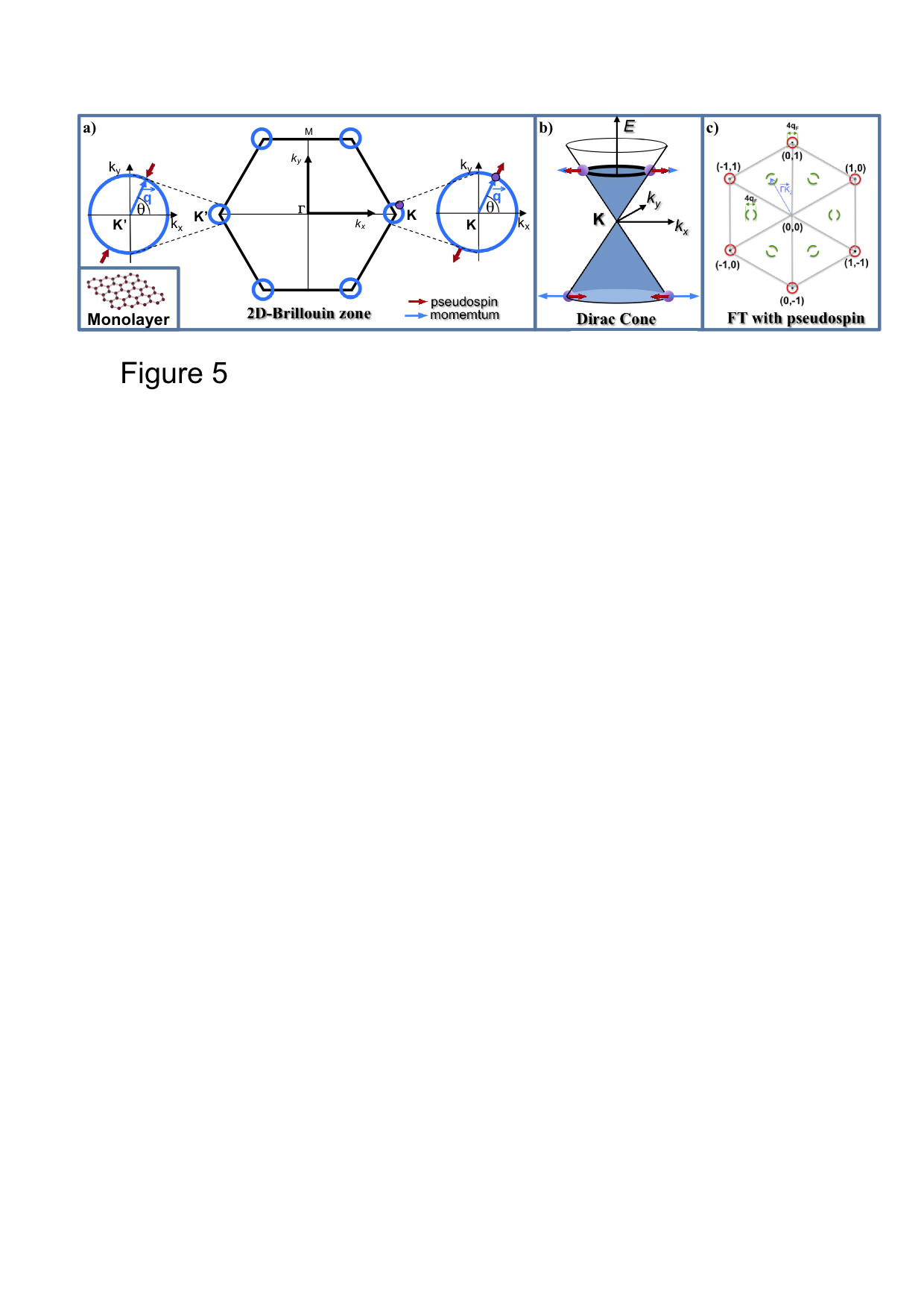}
\end{center}
\label{Figure 1}
\end{figure}

\newpage
\begin{figure}[!h]
\begin{center}
\includegraphics[width=18 cm,clip]{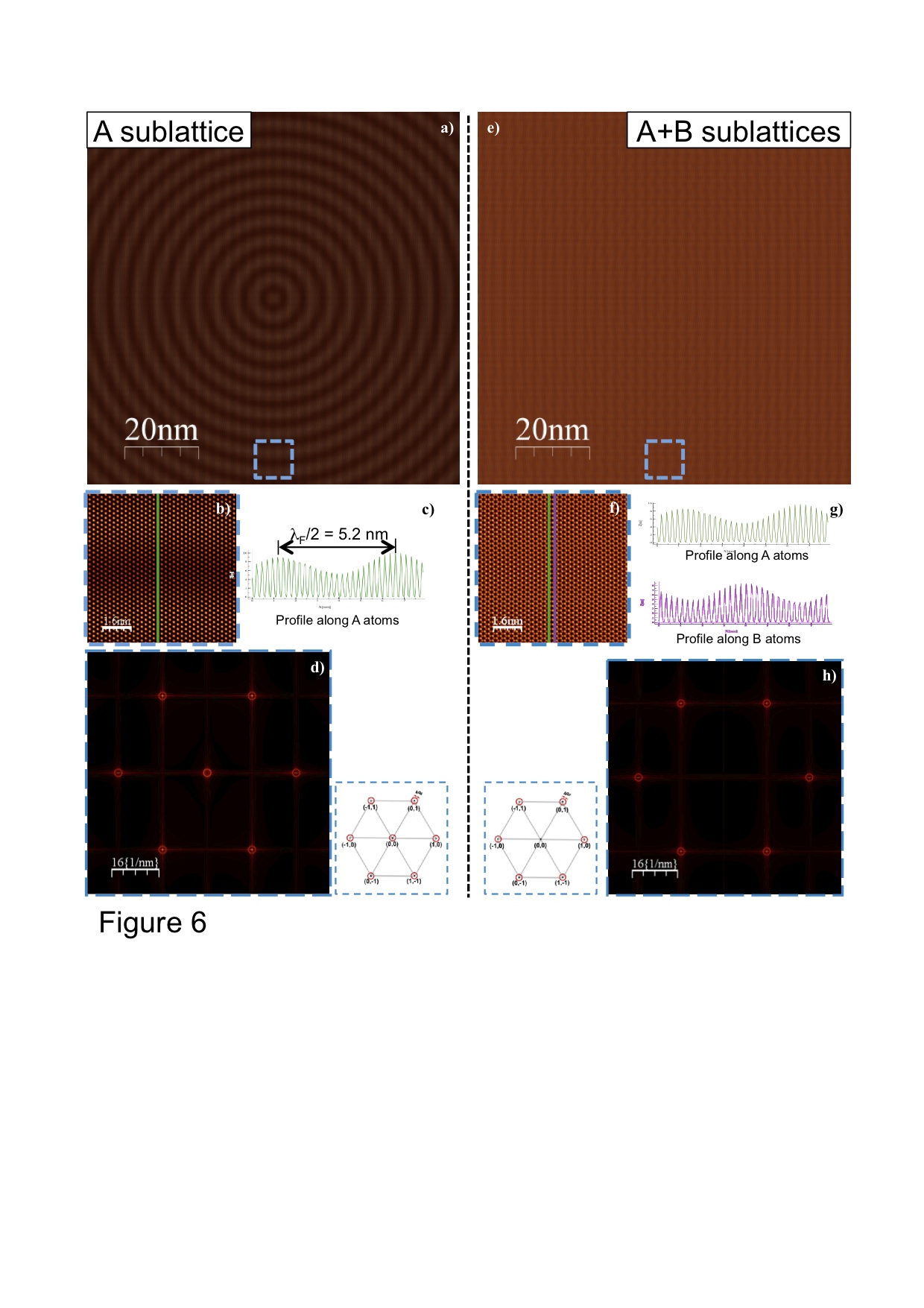}
\end{center}
\label{Figure 1}
\end{figure}

\newpage
\begin{figure}[!h]
\begin{center}
\includegraphics[width=18 cm,clip]{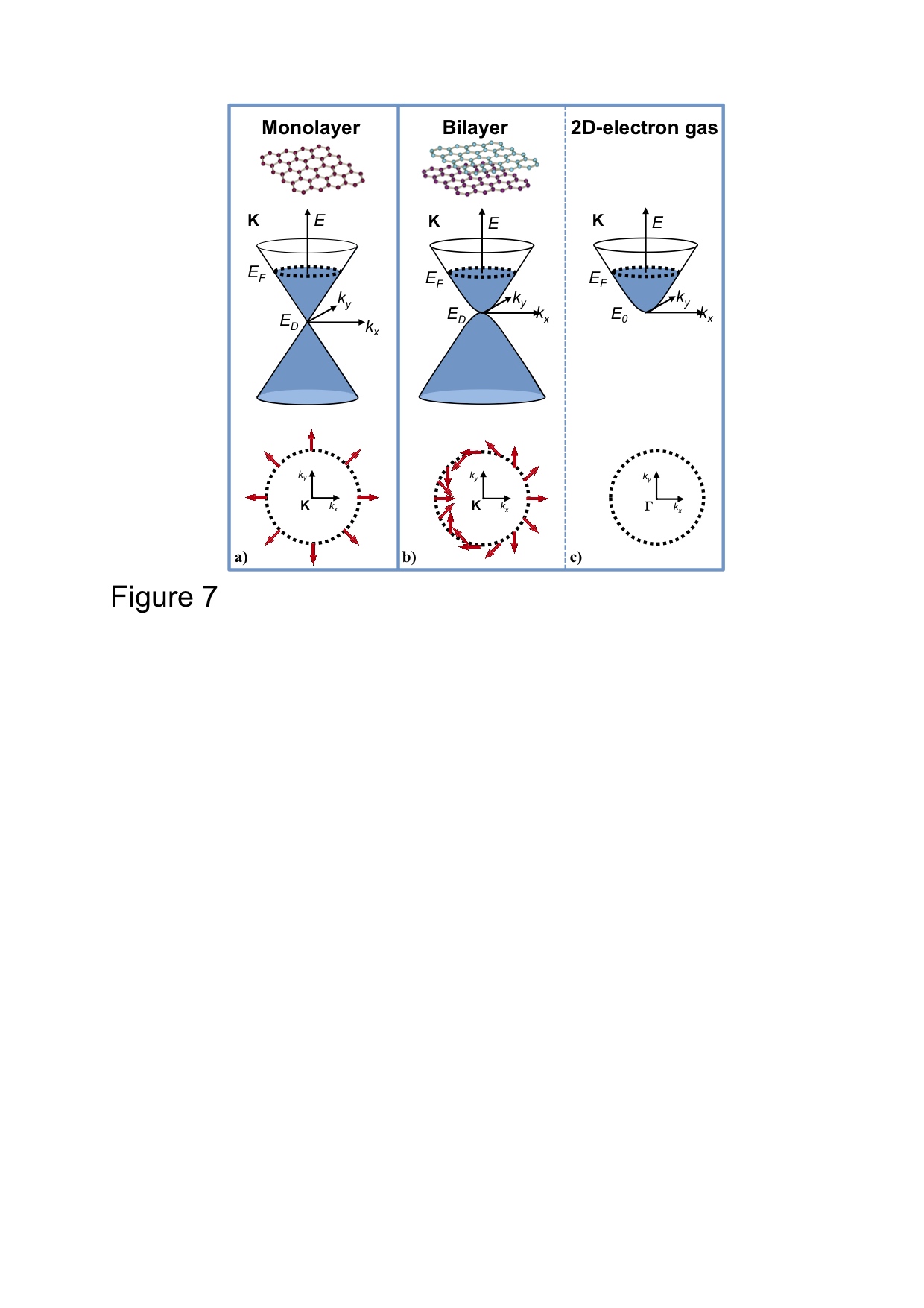}
\end{center}
\label{Figure 1}
\end{figure}

\newpage
\begin{figure}[!h]
\begin{center}
\includegraphics[width=18 cm,clip]{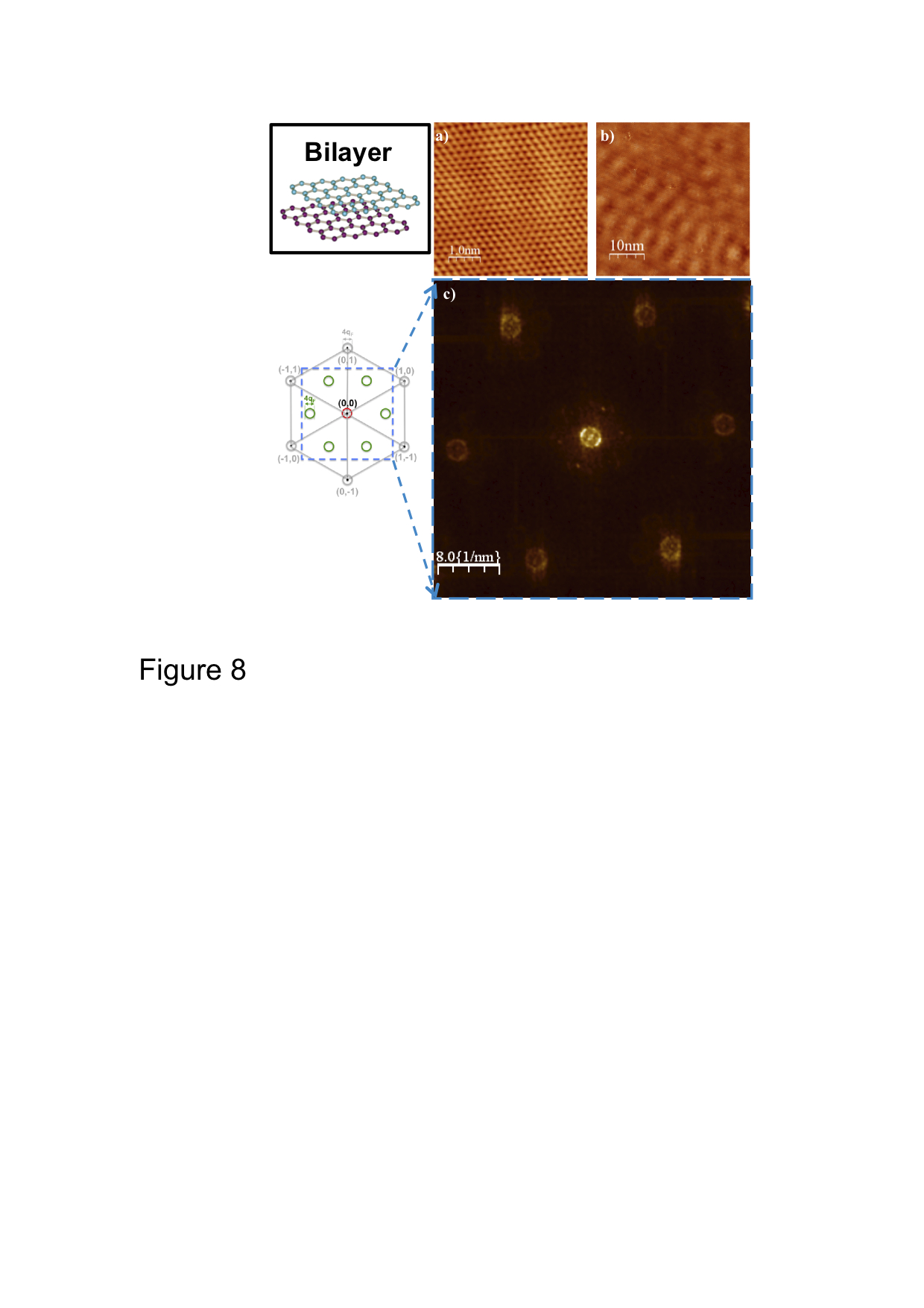}
\end{center}
\label{Figure 1}
\end{figure}

\newpage
\begin{figure}[!h]
\begin{center}
\includegraphics[width=18 cm,clip]{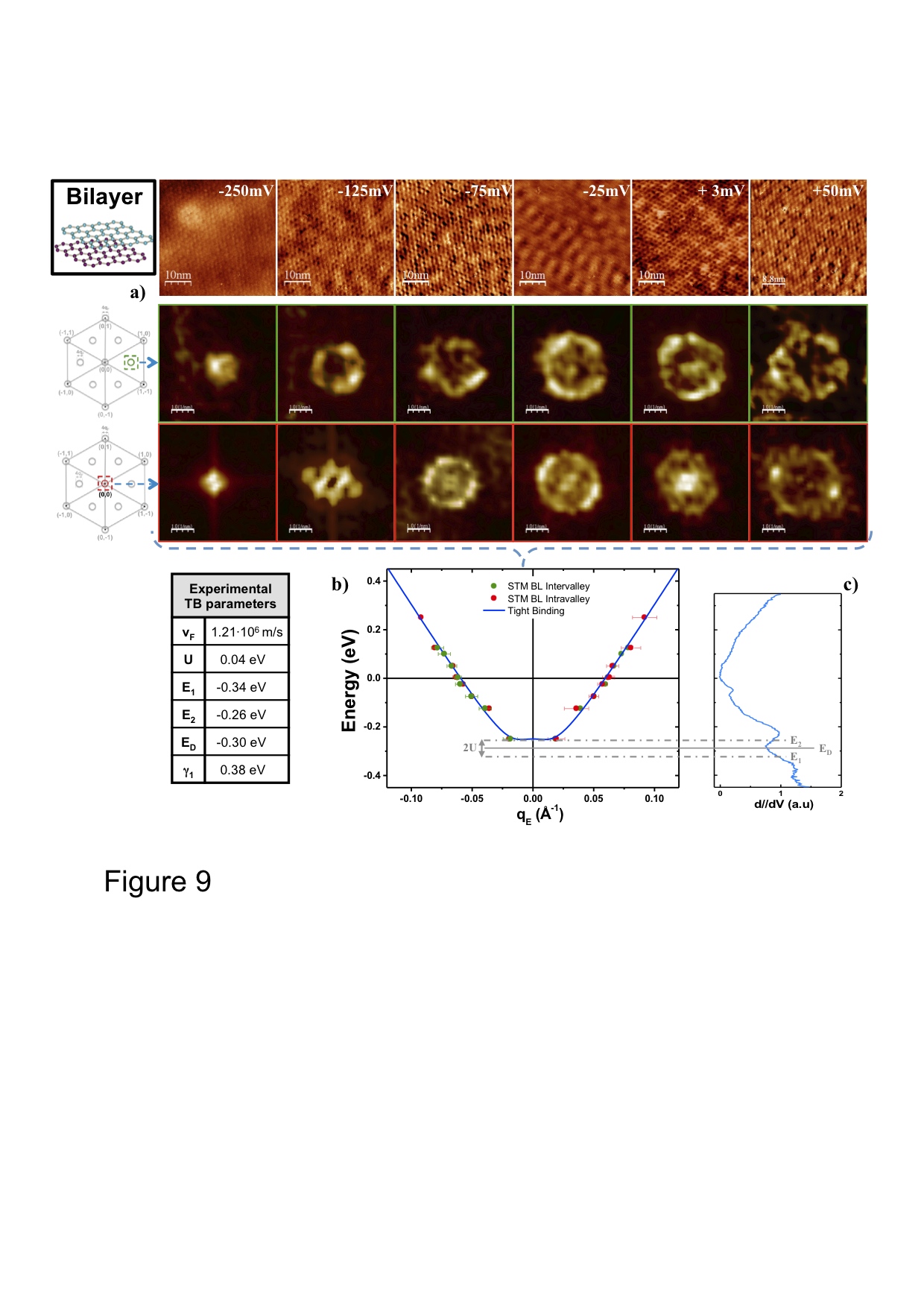}
\end{center}
\label{Figure 1}
\end{figure}

\newpage
\begin{figure}[!h]
\begin{center}
\includegraphics[width=18 cm,clip]{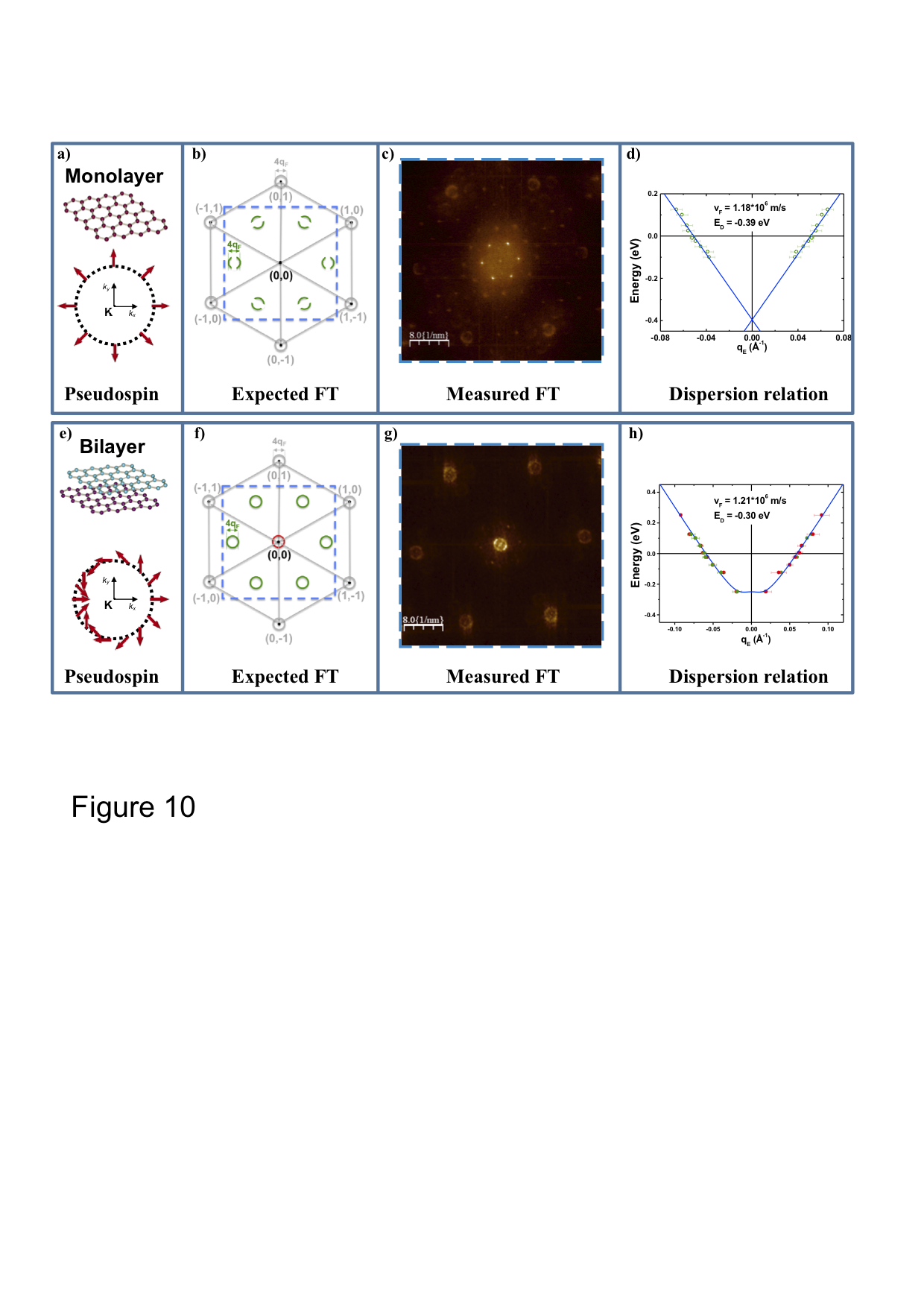}
\end{center}
\label{Figure 1}
\end{figure}


\begin{thebibliography}{99}
\bibitem{Note A}
In the  2D FT-LDOS  maps calculated by Pereg Barnea and Mc Donald in Ref. [19], one should notice that only one sublattice is considered, and thus the central $2q_E$ ring related to intravalley backscattering is restored.
\bibitem{Note B}
 The quantity which is mapped on Fig. 6a  is the function: 


$f_A (x,y) = \left[ {1 + 0.3\cos \left( {2q_F \sqrt {x^2  + y^2 } } \right)} \right]$ 

$\times \left[ {1 + \cos \left( {\sqrt 3 \pi a_G ( - \frac{1}{2}x + \frac{{\sqrt 3 }}{2}y)} \right)} \right] \times \left[ {1 + \cos \left( {\sqrt 3 \pi a_G ( - \frac{1}{2}x - \frac{{\sqrt 3 }}{2}y)} \right)} \right] \times \left[ {1 + \cos \left( {\sqrt 3 \pi a_G x} \right)} \right]$, 



with $q_F$ = 0.6 nm$^{ - 1} $, and $a_G$ = 0.246 nm the graphene lattice parameter. The first term in bracket mimics the $2q_F$ oscillation generated off the impurity located at the center of the image on A site, the remaining terms simulate the triangular $A$ sublattice. A 0.3 arbitrary prefactor is introduced to attenuate the $2q_F$ LDOS modulation with respect to the atomic contrast. Note that the decay of the modulation with distance from the impurity is not taken into account.

The similar quantity $f_B (x,y)$ can be calculated for sublattice $B$ only, by replacing in $f_A$  the $+$ sign by a $-$ sign in the first bracket to inverse the $2q_E$ oscillation, and by replacing $x$ by $x+a$ in the remaining terms to simulate the $B$ sublattice. Finally, Fig. 6e is a 2D representation of $f_A (x,y)+f_B (x,y)$.
\bibitem{Note C} 
The replica rings centered at the first order spots of reciprocal lattice are obtained by the product of the cosine terms appearing in the expression of $f_A$, which lead to beating effects between the the $2q_E$ oscillations and the atomic lattice. This mimics the replica rings expected from scattering of Bloch waves as discussed in section II.
\end{thebibliography}
\end{document}